\documentclass[pra,twocolumn,showpacs,preprintnumbers,amsmath,amssymb,citeautoscript]{revtex4-1}

\usepackage{graphicx}
\usepackage{dcolumn}
\usepackage{bm}
\usepackage{color}
\usepackage{epstopdf} 
\usepackage{enumerate}

\newlength{\figwidth}
\begin{document}

\title{Thermodynamic evidence for nematic phase transition \\ at the onset of pseudogap in YBa$_2$Cu$_3$O$_y$}

\author{Y.~Sato$^1$}
\author{S.~Kasahara$^1$}
\author{H.~Murayama$^1$}
\author{Y.~Kasahara$^1$}
\author{E.-G.~Moon$^2$}
\author{T.~Nishizaki$^3$} 
\author{T.~Loew$^4$}
\author{J.~Porras$^4$}
\author{B.~Keimer$^4$}
\author{T.~Shibauchi$^5$}
\author{Y.~Matsuda$^{1}$}

\affiliation{
$^1$Department of Physics, Kyoto University, Kyoto 606-8502, Japan\\
$^2$Department of Physics, Korea Advanced Institute of Science and Technology, Daejeon 305-701, Korea\\
$^3$Department of Electrical Engineering, Kyushu Sangyo University, Fukuoka 813-8503, Japan\\
$^4$Max Planck Institute for Solid State Research, Heisenbergstra{\ss}e 1, D-70569 Stuttgart, Germany\\
$^5$Department of Advanced Materials Science, University of Tokyo, Chiba 277-8561, Japan\\
}
\date{\today}



\pacs{}

\maketitle
{\bf 
A central issue in the quest to understand the superconductivity in cuprates is the nature and origin of the pseudogap state, which harbours anomalous electronic states such as Fermi arc, charge density wave (CDW), and $\boldsymbol{d}$-wave superconductivity~\cite{Keimer15}. A fundamentally important, but long-standing controversial problem has been whether the pseudogap state is a distinct thermodynamic phase characterized by broken symmetries below the onset temperature $\boldsymbol{T^*}$.  Electronic nematicity, a fourfold ($\boldsymbol{C_4}$) rotational symmetry breaking, has emerged  as a key feature inside the pseudogap regime~\cite{Kivelson98,Vojta09,Ando02,Hinkov04}, but the presence or absence of a nematic phase transition and its relationship to the pseudogap remain unresolved.  Here we report thermodynamic measurements of magnetic torque in the underdoped regime of orthorhombic YBa$\boldsymbol{_2}$Cu$\boldsymbol{_3}$O$\boldsymbol{_y}$ with a field rotating in the CuO$\boldsymbol{_2}$ plane, which allow us to quantify magnetic anisotropy with exceptionally high precision.  Upon entering the pseudogap regime, the in-plane anisotropy of magnetic susceptibility increases after exhibiting a distinct kink at $\boldsymbol{T^*}$.  
Our doping dependence analysis reveals that this anisotropy is preserved below $\boldsymbol{T^*}$ even in the limit where the effect of orthorhombicity is eliminated.  In addition, the excess in-plane anisotropy data show a remarkable scaling behaviour with respect to $\boldsymbol{T/T^*}$ in a wide doping range.  
These results provide thermodynamic evidence that the pseudogap onset is associated with a second-order nematic phase transition, which is distinct from the CDW transition that accompanies translational symmetry breaking~\cite{Ghiringhelli12, Chang12,Comin15b,Hucker14,Blanco-Canosa14,Wu15,Jang16,Sebastian15}. This suggests that nematic fluctuations near the pseudogap phase boundary have a potential link to the strange metallic behaviour in the normal state, out of which high-$\boldsymbol{T_c}$ superconductivity emerges. %
}

Nematicity has been widely discussed in cuprates and one of its mechanism is the onset of a stripe-type CDW order parameter which generally breaks rotation symmetry as well as translation symmetry with a nonzero wave number $\bm{Q}\neq0$~\cite{Ghiringhelli12, Chang12,Comin15b,Hucker14,Blanco-Canosa14,Wu15,Jang16,Sebastian15,Scalapino,Wang14,Schutt15,Yamakawa15}.  
In Bi$_2$Sr$_2$CaCu$_2$O$_{8+\delta}$ (BSCCO), the scanning tunneling microscope (STM) experiments at low temperatures report an electronic state, consisting of short-range CDW of unidirectional (one-dimensional, 1D) type with the period of $\sim4a_0$, where $a_0$ is the Cu-O-Cu distance~\cite{Kohsaka07,Lawler10}.  This nano-stripe structure persists even well above the superconducting transition temperature $T_c$~\cite{Parker10}. In YBa$_2$Cu$_3$O$_y$ (YBCO), the short-range CDW order forms a dome-shaped boundary inside the pseudogap regime~\cite{Hucker14, Blanco-Canosa14}. Resonant X-ray scattering (RXS) experiments  in YBCO report that the CDW is of unidirectional type with the periodicity of  $\sim3a_0$~\cite{Comin15b}.  In both BSCCO and YBCO, the CDW forms domains with the size of $\sim 3$\,nm in zero field, inside which  the $C_4$ symmetry of the unit cell is strongly broken.  In contrast to such CDW orders, the nematicity may also be caused by an instability without breaking translational symmetry, characterized by $\bm{Q}=0$.  

The measurement of the magnetic torque has a very high sensitivity for detecting magnetic anisotropy.
The torque {\boldmath $\tau$}=$\mu_0V${\boldmath $M$}$\times$ {\boldmath $H$} is a thermodynamic quantity, a differential of the free energy with respect to angular displacement.  Here $\mu_0$ is the permeability of vacuum, $V$ is the sample volume, and {\boldmath $M$} is the magnetization induced by the external magnetic field {\boldmath $H$}.  When {\boldmath $H$} is rotated within the $ab$-plane, {\boldmath $\tau$} is a periodic function of double the azimuthal angle $\phi$ measured from the $a$ axis (Fig.\,1a):
\begin{equation}
\tau_{2\phi}=\frac{1}{2}\mu_0H^2[(\chi_{aa}-\chi_{bb})\sin2\phi-2\chi_{ab}\cos2\phi],
\end{equation}
where the susceptibility tensor $\chi_{ij}$ is given by $M_i=\Sigma_j\chi_{ij}H_j$. In a system with a tetragonal symmetry, $\tau_{2\phi}$ should be zero.  When $C_4$ symmetry is broken, a nonzero $\tau_{2\phi}$ appears as a result of $\chi_{aa}\neq \chi_{bb}$ and/or $\chi_{ab}\neq0$ depending on the orthorhombicity direction.

\begin{figure}[h!]
\begin{center}
\includegraphics[width=1.0\linewidth]{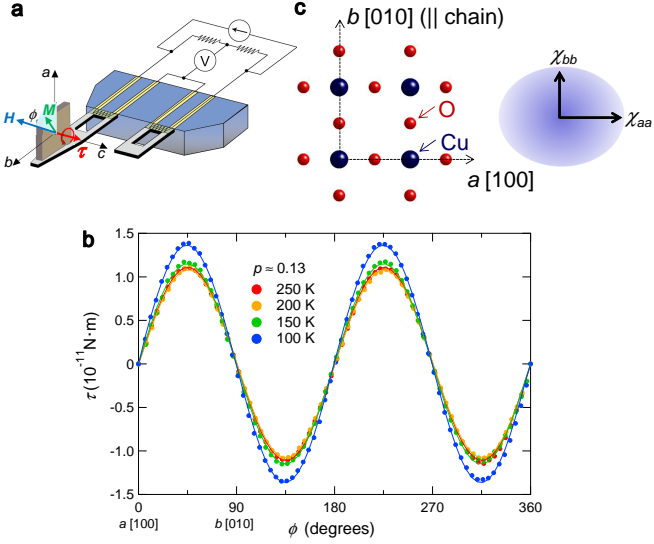}
\end{center}
\caption{
{\bf In-plane torque magnetometry in YBCO single crystals.} 
{\bf a,} Schematic representations of the experimental configuration for torque measurements under in-plane field rotation. An untwinned single-crystalline sample of YBCO is mounted on the piezo-resistive lever which forms an electrical bridge circuit with the neighbouring reference lever. 
{\bf b,} Typical curves of magnetic torque $\tau$ in YBCO ($p\sim0.13$) as a function of the azimuthal angle $\phi$ from the $a$ axis. 
The field of $\mu_0H = 7$\,T is applied within the $ab$ plane with the misalignment less than 0.1$^{\circ}$. 
{\bf c,} Schematic view of the CuO$_2$ plane. Because of the orthorhombic crystal structure stabilized by CuO chains ($\parallel$ $b$ axis ) lying between the CuO$_2$ bilayers, $\chi_{aa}>\chi_{bb}$. 
} 

\end{figure}

\begin{figure}[t]
\begin{center}
\includegraphics[width=1.0\linewidth]{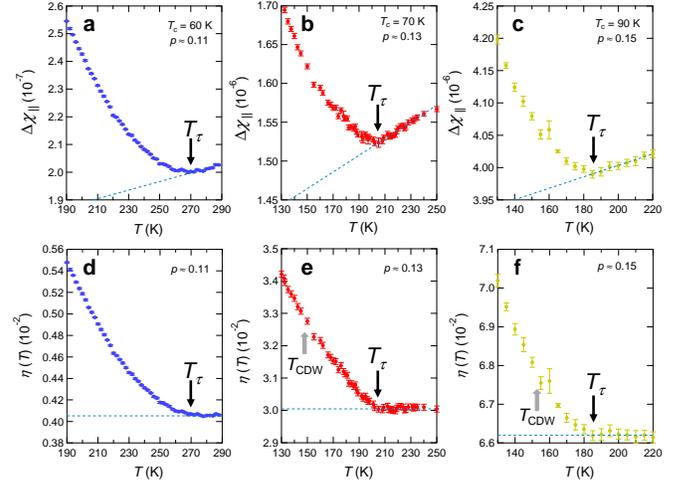}
\end{center}
\vfill \caption{
{\bf In-plane anisotropy of the magnetic susceptibility for underdoped YBCO with different hole doping levels.}
{\bf a, b, c,} Temperature dependence of the in-plane anisotropy of the magnetic susceptibility,  $\Delta\chi_\parallel = \chi_{aa}-\chi_{bb}$, determined from the torque curves for underdoped YBCO with hole concentration $p \approx 0.11, 0.13$, and 0.15.  For all crystals, $\Delta\chi_\parallel$  gradually decreases down to $T_\tau$ and increases rapidly below $T_\tau$ after exhibiting a kink at $T_\tau$. 
{\bf d, e, f,} Temperature dependence of the order parameter $\eta \equiv (\chi_{aa}-\chi_{bb})/(\chi_{aa}+\chi_{bb})$ for $p \approx 0.11, 0.13$ and 0.15. 
Above $T_\tau$,  $\eta(T)$ is temperature independent  while it shows a steep increases below $T_\tau$.   In sharp contrast to the anomaly at $T_\tau$, no discernible anomaly is observed at CDW transition temperature $T_{\rm CDW}$.   The background anisotropy due to the crystal orthorhombicity, $\eta(T_\tau)$, increases rapidly with $p$. 
 } 
\end{figure}

\begin{figure}[t]
\begin{center}
\includegraphics[width=0.9\linewidth]{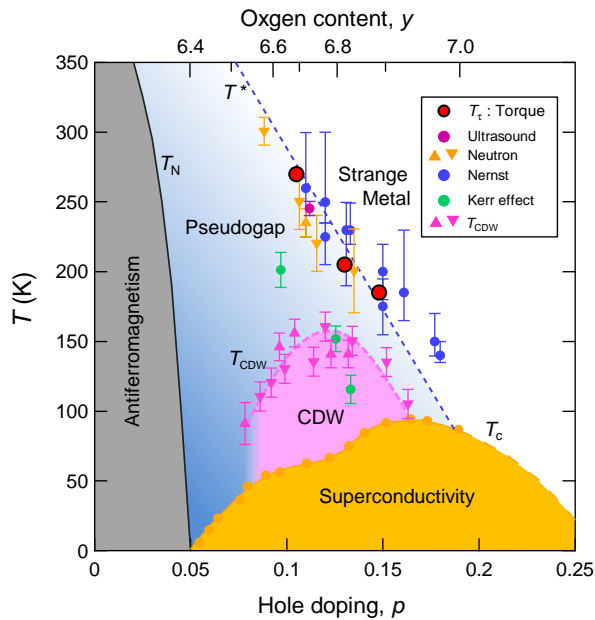}
\end{center}
\caption{{\bf Temperature-doping phase diagram of YBCO.} 
The phase diagram contains at least four different ordered phases, including antiferromagnetism (gray), superconductivity (yellow), CDW (pink) and pseudogap (blue) regimes.   
Pseudogap line (dashed line) at $T^*$ marks the boundary between the strange metal and even more anomalous regimes.  
Red circles represent the second-order nematic transition temperature $T_{\tau}$ determined by the present in-plane torque magnetometry.  For comparison, the pseudogap temperatures determined by other probes are also plotted. 
Purple circles, orange triangles and blue circles are $T^*$ reported by ultrasound spectroscopy~\cite{Shekhter13},  polarized neutron scattering~\cite{Fauque06,
Mook08}, and Nernst coefficient~\cite{Daou10}, respectively.  
Magenta triangles represent the formation temperature of the short range CDW, $T_{\rm CDW}$, reported by resonant X-ray measurements~\cite{Hucker14,Blanco-Canosa14}.  
Green circles are the temperature below which the time reversal symmetry is broken, 
reported by the polar Kerr effect~\cite{Xia08}. 
}

 \label{Fig1}
\end{figure}

\begin{figure}[t]
\begin{center}
\includegraphics[width=1.0\linewidth]{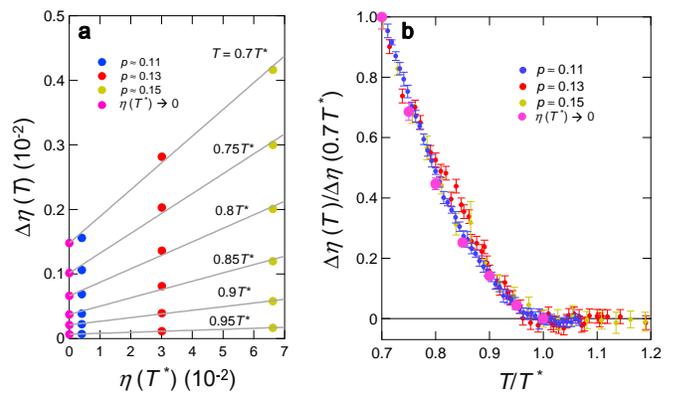}
\end{center}
\caption{
{\bf Induced nematicity and the scaling behaviour.}
{\bf a,} The excess anisotropy below $T^*$, $\Delta \eta(T) \equiv \eta(T)-\eta(T^*)$, of YBCO with different hole concentrations $p$, plotted as a function of the background anisotropy $\eta(T^*)$ at different values of $T/T^*$.  The solid lines represent the linear fit for $\Delta \eta(T)$ at each $T/T^*$. The magenta symbols show the induced nematicity in the limit of  $\eta(T^*)\rightarrow0$. 
{\bf b,} The excess anisotropy $\Delta \eta(T)$ normalized by the value at $T/T^*=0.7$ plotted as a function of $T/T^*$.  All the data collapse into a universal curve, indicating a scaling relation. 
} 
\end{figure}

Figure\,1b displays typical curves of magnetic torque measured as a function of $\phi$. All the curves are perfectly sinusoidal. 
In the temperature range shown here, oscillations are proportional to $\sin 2\phi$ with positive sign, indicating  $\chi_{aa} >\chi_{bb}$ and $\chi_{ab}=0$ (Fig.\,1c).   Figures 2a, b, and c depict the amplitude of in-plane anisotropy of the susceptibility,  $\Delta \chi_{\parallel}\equiv \chi_{aa}-\chi_{bb}$,  at $\mu_0H=7$\,T for underdoped YBCO with $T_c=60$, 70, and 90\,K (hole concentration $p \approx 0.11$, 0.13, and 0.15), respectively.  In all the crystals, as the temperature is lowered, $\Delta \chi_{\parallel}$ gradually decreases and increases rapidly after exhibiting a distinct kink at $T=T_{\tau}$.  Since the average of uniform susceptibilities $\chi_{aa}$ and $\chi_{bb}$ is also temperature dependent, we introduce a dimensionless order parameter  $\eta\equiv  (\chi_{aa}-\chi_{bb})/(\chi_{aa}+\chi_{bb})$, a diagonal component of a nematic traceless symmetric tensor in two spatial dimensions,  to discuss the nematicity properly (see Figs.\,2d, e, f). 
Above $T_{\tau}$, $\eta(T)$ is nearly temperature independent, indicating that the uniform susceptibility causes the weak temperature dependence of $\Delta \chi_{\parallel}$ above $T_{\tau}$.   Below $T_{\tau}$, $\eta(T)$ increases with a slightly concave curvature.  

Figure\,3 displays the temperature-doping phase diagram of YBCO.  Obviously $T_{\tau}$  coincides well with the pseudogap temperature $T^*$ determined by various other probes.  In what follows, we identify  $T_{\tau}$ as the pseudogap onset temperature $T^*$, i.e. $T_{\tau}=T^*$.   The kink anomaly in the temperature dependence of $\Delta\chi_{\parallel}$  is usually an indication of  a second-order phase transition.  However, the $C_4$ symmetry is already broken due to the orthorhombic crystal structure with 1D CuO chains in YBCO, and thus no further rotational symmetry breaking is expected. 
Indeed, $\eta(T)$ is finite even above $T^*$ for all the samples (Figs.\,2d, e, and f), confirming that the orthorhombic structure generally leads to anisotropic magnetic susceptibility. We note that the magnitude of $\eta(T^*)$ increases with hole doping, which may be explained by the increased crystal orthorhombicity through the oxidization of CuO chains with doping. 

Disentangling an intrinsic electronic nematicity in the CuO$_2$ planes from extrinsic effects due to crystal orthorhombicity has been a vexing issue particularly in YBCO. It should be stressed that the doping dependence of $\eta(T)$ enables us to examine the nematicity in the limit where the effect of orthorhombicity is removed. Since $\eta(T)$ is temperature independent above $T^*$, $\eta(T^*)$ represents the background anisotropy stemmed from the crystal orthorhombicity. To eliminate this background contribution, we introduce the excess nematicity below $T^*$,  $\Delta \eta(T)\equiv  \eta(T) - \eta(T^*)$, and plot it as a function of background anisotropy $\eta(T^*)$ (Fig.\,4a).   As shown by the solid lines,  $\Delta \eta(T)$ is nearly proportional to $\eta(T^*)$. 
Obviously the solid lines have finite intercepts, indicating that even when the crystal orthorhombicity is removed, nematicity remains finite below $T^*$; i.e. spontaneous $C_4$ rotational symmetry breaking in the pseudogap state. This result, along with the kink anomaly of the in-plane torque (Figs.\,2a, b, and c), provides evidence for a second-order phase transition at $T^*$ in the CuO$_2$ planes of YBCO. 

The second-order nematic phase transition at $T^*$ is further supported by the scaling property of the nematicity for crystals with various hole concentrations.   Although $T^*$ and $\eta(T^*)$ have both strong doping dependence, the excess anisotropy $\Delta \eta(T)$ exhibits a good scaling behaviour when plotting as a function of $T/T^*$.  Figure\,4b depicts  $\Delta \eta(T)$ normalized by the value at $T=0.7 T^*$ vs. $T/T^*$.  All the curves collapse into a single curve in a wide temperature range.  Moreover, the data in the limit of no background anisotropy (Fig.\,4a) lie on the same curve.  
It is well known that the genuine second order phase transitions do not occur in the presence of external symmetry-breaking field and the kink-type temperature dependence of order parameters will be smeared out. However, if the external field is small enough, the kink behaviours are only modified slightly near the transition points, and scaling properties should prevail.
 In the present case, the crystal orthorhombicity in YBCO is the external symmetry-breaking field. At the same time, the orthorhombic distortion would be helpful to prevent the formation of microscopic domains with orthogonal nematic directions, and thus quite important for the twofold nematic signals to be observed in the bulk measurements \cite{Daou10}. The data collapse into the universal curve in Fig.\,4b indicates that the influence of the background anisotropy on the nematic order parameter is small except in the vicinity of $T^*$. This supports that the crystal orthorhombicity is a small perturbation on the second-order transition and reinforces our analysis of background subtraction.
 The characteristic super-linear temperature dependence of $\Delta \eta(T)$ appears at the onset of the nematicity. If one interprets the temperature dependence with the critical exponent $\beta$, $\Delta \eta(T)\propto(T^*-T)^\beta$, then the critical exponent shows large deviations from all known results of two dimensional nematic transitions, for example ones of mean-field ($\beta=1/2$) and the 2D Ising model ($\beta=1/8$).  Thus, the super-linear dependence indicates the nematic transition at $T^*$ is in a very different universality class and calls for further theoretical investigation including scenarios of composite order parameters with randomness and doped spin liquids \cite{Nie2014,Sachdev}

Although the second-order phase transition at $T^*$ has been suggested by several experiments, it is far from settled. Resonant ultrasound spectroscopy experiments report the critical slowing down behaviour in the ultrasound absorption as $T^*$ is approached~\cite{Shekhter13}, but a different interpretation without critical phenomena has been proposed~\cite{Cooper14}. The polarized neutron scattering experiments report the time reversal symmetry breaking (TRSB) with appearance of magnetic moment at $T^*$, which has been interpreted as the circulating current loops within the CuO$_2$ unit cell~\cite{Fauque06,
Mook08}. However, the polar Kerr effect measurements report that the TRSB temperature is significantly different from $T^*$~\cite{Xia08}. The enhancement of in-plane anisotropy of the Nernst coefficient at $T^*$ has been reported~\cite{Daou10}, but more recent results have shown that such an enhancement is much more pronounced below $T_{\rm CDW}$ rather than $T^*$~\cite{Cyr-Choiniere15}. These results are in sharp contrast to our torque experiments in which no discernible anomalies are observed at $T_{\rm CDW}$. Recent RXS experiments report the appearance of orthogonal CDW domains with $\bm{Q}_1 \approx(1/3,0,L)$ and $\bm{Q}_2 \approx(0,1/3,L)$~\cite{Comin15b}. Our results suggest that the effective cancellation of the nematicity occurs due to nearly equal numbers of these CDW domains. We also point out that no anomaly in $\eta(T)$ is expected at $T_{\rm CDW}$ when the CDW is of bidirectional type (checkerboard) \cite{Wang14,Sebastian15}, which preserves rotational symmetry,  is formed.

Our results indicate that the pseudogap state is an electronic nematic phase. The phase diagram of hole-dope cuprate superconductors (Fig.\,3) then include at least 4 different ordered phases; antiferromagnetic, superconducting, CDW, and pseudogap phases, which are characterized by broken time, gauge, translational, and rotational symmetries, respectively. 
No anomalies with $\bm{Q}\neq0$ have been reported by various types of diffraction measurements at $T^*$.  Therefore the observed nematic transition at the pseudogap line is most likely to be attributed to a ferro-type instability with $\bm{Q}=0$.

Angle-resolved photoemission spectroscopy (ARPES) experiments in BSCCO and related compounds revealed the Fermi arc where the Fermi surface is partially disappeared in the pseudogap state~\cite{Hashimoto14}. Yet, important questions still remain, for instance, the link between the nematic transition and Fermi arc formation and the interplay between pseudogap and CDW.  
Whether a quantum critical point (QCP) is present inside the superconducting dome has been a hotly debated issue in cuprates~\cite{Keimer15, Shekhter13, Cooper14}.  
The presence of QCP has been suggested by the quantum oscillations and ARPES measurements at around $p\approx0.18$ - 0.20~\cite{Sebastian15,Hashimoto14}. 
The identification of the pseudogap temperature as the critical temperature of a second-order $\bm{Q}=0$ nematic transition favours the QCP scenario; i.e. the extension of the pseudogap temperature to $T\rightarrow0$ suggests a nematic QCP.  The second-order nature of the phase transition line, in general, implies the presence of critical fluctuations near the transition line, and in an extended regime around the QCP one may expect significant quantum critical fluctuations. 
Hence it is tempting to consider that the nematic quantum fluctuations influence the superconductivity as well as the strange metallic behaviour in the normal state of cuprates.

\section*{Methods}

\noindent {\bf Materials.}
High-quality single crystals of YBCO were grown by the self-flux method using a Y$_2$O$_3$ crucible~\cite{Naito97}. In the present study we used naturally untwinned single crystals which were carefully selected under a polarized microscope. The oxygen concentration was controlled by annealing the crystals at high temperatures under oxygen or nitrogen flow atmosphere~\cite{Nishizaki08}. 
The superconducting transition temperature $T_c$ was characterized by the magnetization measurements. The crystals exhibit sharp superconducting transitions~\cite{Nishizaki08} with the $T_c$ determined as the midpoint at 60, 70, and 90 K for $p \approx 0.11, 0.13,$ and 0.15, respectively. 
First-order vortex-lattice melting transition, which can be seen only in clean and homogeneous single crystals, is clearly observed in the crystals prepared by the same method~\cite{Nishizaki99,Nishizaki00a,Nishizaki00b}, indicating the high quality of our crystals. 
For each crystal, the directions of the $a$ and $b$ axes were determined by X-ray diffraction.  

\bigskip

\noindent {\bf Torque magnetometry.} 
Magnetic torque is measured by the piezo-resistive micro-cantilever technique, which is a very sensitive probe of magnetic anisotropy~\cite{Okazaki11,Kasahara12,Watanabe12,Kasahara16}. 
In this method, an isotropic Curie contribution from impurity spins is cancelled out~\cite{Watanabe12}. 
Carefully selected single crystals with typical dimensions of 250$\times$250$\times$50 $\mu$m$^3$ are used in the torque measurements. The in-plane and out-of-plane anisotropies of the magnetic susceptibilities can be measured depending on the geometry of the sample, which is mounted on the lever.

\section*{Acknowledgements} 
We thank A. Carrington, R.\,M. Fernandes, T. Hanaguri, N. Harrison, S.\,M. Hayden, M.-H. Julien, S. Kivelson, H. Kontani, C. Putzke, T.\,M. Rice, S. Sachdev, L. Taillefer, T. Tohyama, H. Yamase, and J. Zaanen for fruitful discussions, and M. Ishikawa and H. Yamochi for experimental support. This work was supported by Grants-in-Aid for Scientific Research (KAKENHI) (Nos. 25220710, 15H02106, 15H03688, 16K05460, 16K13837) and on Innovative Areas ``Topological Material Science'' (No. 15H05852) from Japan Society for the Promotion of Science (JSPS). The characterization of YBCO single crystals was partly performed at Advanced Instruments Center at Kyushu Sangyo University. E.-G. M. acknowledges the financial supports from the POSCO Science Fellowship of POSCO TJ Park Foundation and NRF of Korea under Grant No. 2017R1C1B2009176.

\section*{\bf Author contributions} 
T.N., T.L., J.P. and B.K. prepared the high-quality single crystalline samples. Y.S., H.M. and S.K. performed the magnetic torque measurements. Y.S., S.K., E.-G.M. and Y.M. analysed the data. S.K., E.-G.M, Y.K., T.S., B.K. and Y.M. discussed and interpreted the results and prepared the manuscript.

\end{document}